\documentclass[twocolumn,aps,pra,showpacs,amsmath,superscriptaddress]{revtex4}
\usepackage{amssymb}
\usepackage{mathrsfs}
\usepackage{graphicx}
\usepackage{float}

\begin{document}

\title{Topological insulator states in quasi-one-dimensional sawtooth chain}

\author{Zhihao Xu}
\email{xuzhihao@outlook.com}
\affiliation{Beijing National
Laboratory for Condensed Matter Physics, Institute of Physics,
Chinese Academy of Sciences, Beijing 100190, China}
\affiliation{Department of Physics and Center of Theoretical and Computational Physics,
The University of Hong Kong, Hong Kong, China}

\author{Linhu Li}\affiliation{Beijing National
Laboratory for Condensed Matter Physics, Institute of Physics,
Chinese Academy of Sciences, Beijing 100190, China}

\date{ \today}

\begin{abstract}
We report the discovery of flatten Bloch bands with nontrivial topological numbers in a
quasi-one-dimensional sawtooth chain. We present the nearly flat-band with a obvious gap and nonzero
Chern number of the modulated sawtooth chain by tuning hoppings and modulation. With the increasing of the strength of the modulation,
the system undergos a series of topological phase transitions in the absence of interaction. By adding interaction to the model,
we present exact diagonalization results for the system at $1/3$ filling and some interesting connections with
fraction quantum Hall states: topological degeneracy, nontrivially topological number and fractional statistic of the quasihole.
\end{abstract}


\maketitle

\section{Introduction}
The studying of fractional quantum Hall effect(FQHE)\cite{Tsui}
brings about a new upsurge in condensed matter physics in the past decades.
Many theories try to understand the nature of the FQHE found in two-dimensional($2$D)
systems with a strong magnetic fields by introducing fractionally changed
quasi-particles\cite{Laughlin,Haldane,Halperin}, flux attachment and composite
Fermi-liquid theory\cite{Jain}. And the properties of fractional topological states
with different geometrical configurations are extensive studied\cite{Haldane,Kitaev,Seidel,Bergholtz-PRL,Bergholtz2,DHLee}.
However, the cases in lattice models are still unclear. Many efforts in discovering the quantum Hall effect(QHE)
in various kinds lattices in recent years. Haldane\cite{Haldane1} proposes his honeycomb lattices with
the breaking time-reversal invariant, which is the milestone in finding QHE in lattices. Another breakthrough
is the finding of a new class of topological insulator$-$$Z_2$topological insulator\cite{Kane1,Kane2,SCZhang}. Due to the extended band,
both systems do not reach fractional topological regime where ground states topological degeneracy and
fractional excitations can be found.

The pivotal purpose is to find the lattice models with topologically nontrivial flat-bands which is
similar to Landau-levels and it has drawn great interests in the recent few years. Since $2011$,
a class of topologically nontrivial flat-band models are presented\cite{Kai,Tang,Neupert}
and the ratios of the band gap to the bandwidth reach $20-50$.
By considering the repulsive interaction\cite{Scaffdi,Regnault,Yang,Wang,Yang1,Liu,Sheng,Regnault1,Wang1,Zhao},
fractional Chern topological insulator has been found and the feature of such insulator are
fractionalization and topological degeneracy which is also the key of the traditional FQHE\cite{Wen}.
While most of such systems are focus on $2$D case until recently. One-dimensional($1$D) fermi gas
loaded in a bichromatic optical lattice with commensurate or incommensurate modulations\cite{Lang,Kraus,LangPRB,LangArxiv,xuzhihao,XZH,Haiping},
due to adding another additional dimension, they find such systems have a nonzero Chern number.
Although the energy bands in the Letter\cite{Lang} are still highly dispersive, is it possible to
construct a topologically nontrivial flat-band model by adding a periodic modulation to an
flat-band system in $1$D and form a insulator with the properties similar to FQHE when
interactions are taken into account?

One of the simplest models with a flat-band is fermions loaded in a standard sawtooth chain for the hopping parameters
$t_2=\sqrt{2}t_1$, where $t_1$ is the hopping amplitude along the baseline and
$t_2$ is the one along the diagonal line.
The bottom band is flat but topologically trivial. In this paper, we present the
commensurate modulating sawtooth chain with nearly flat-band parameters with nontrivially topological
properties. For a periodic case, the number of single-particle energy bands depends on the period of
the modulation. Each band has different nontrivial Chern numbers and with the changing of the ratio of
the modulation strength and the hopping amplitude, the topological phase transitions emerge which are distinguished
by the changing of topological numbers and the gap closing.  For open
boundary condition(OBC), the edge states are found in the gaps, and the edge modes connect the neighbouring
energy bands in the gaps with the rolling the phase of the modulation, which indicates the bulk states
are topologically nontrivial. We further display the exact diagonalization results of the one with $1/r$-tails
interaction. We discover fractional topological states at filling factor $1/3$ in such model which is
characterized by the topological quasidegenerate ground states in finite-size systems and fractional excitations.
The lowest quasidegenerate states possess a unit total Chern number and the topological properties of the system
are similar to the usual FQHE on the torus which can establishing relationship between the flux quanta and the
momentum of the fractional topological insulator in $1$D.

We organize our paper as follows. In Sec.II, we introduce the model Hamiltionian,
which is the fermions loaded in the modulated sawtooth chain with
long-range interactions. In Sec. III, we first consider the noninteraction parts.
The band structure, edge states and phase transitions of such model are presented.
Then, the exact diagonalization method is used to discuss the fractional
topological properties of our model with long-range interactions at filling factor $1/3$.
Finally, the summary is in Sec. IV.

\section{Model Hamiltonian}
We consider fermions loaded in the the sawtooth chain with a periodic modulation,
which is described by $H=H_0+H_I$ with
\begin{eqnarray}
\label{eqn1}
H_0&=&\sum_{i}{[t_1\hat{c}_{2i}^{\dag}\hat{c}_{2i+2}+t_2(\hat{c}_{2i}^{\dag}\hat{c}_{2i+1}+\hat{c}_{2i+1}^{\dag}\hat{c}_{2i+2})+\mathrm{h.c.}]} \nonumber \\
& & +\sum_i{\lambda_i \hat{n}_i},
\end{eqnarray}
and
\begin{equation}
\label{eqn2}
H_I=\sum_{i\neq j}{\frac{\hat{n}_i\hat{n}_j}{|\vec{r}_i-\vec{r}_j|}},
\end{equation}
where $\hat{c}^{\dag}_{i}$($\hat{c}_{i}$) is the creation(annihilation)
operator of the fermion, $\hat{n}_{i}=\hat{c}^{\dag}_{i}\hat{c}_{i}$ is the particle number operator,
$t_1$ is the hopping strength along the baseline which is set to
the unit and $t_2$ is the one along the diagonal directions. For standard sawtooth chain for fermions under periodic
boundary condition(PBC), two energy bands are $\varepsilon_{\pm}(k)=t_1\cos{k}\pm \sqrt{t_1^2\cos^2{k}+2t_2^2(1+\cos{k})}$.
Setting $t_2=\sqrt{2}t_1$, the bottom band becomes completely flat, i.e. $\varepsilon_{-}(k)=-2t_1$,
and the the dispersion band is $\varepsilon_{+}(k)=2t_1(1+\cos{k})$ with the energy gap $\Delta=2t_1$. The
dispersionless band constructs localized excitations. The fermion is localized in any of the $L/2$
valleys labeled by the index $2i$ with energy $\varepsilon_{-}(k)$ given by $l_{2i}^{\dag}|0\rangle$ where
$l_{2i}^{\dag}=(1/2)(\hat{c}_{2i-1}^{\dag}-\sqrt{2}\hat{c}_{2i}^{\dag}+\hat{c}_{2i+1}^{\dag})$\cite{Richter},
where $L$ is lattice number.
$\lambda_i=\lambda\cos(2\pi\alpha i+\delta)$, where $\lambda$ is the modulation
strength of the commensurate potential, $\delta$ is an arbitrary phase offering another degree of freedom.
For commensurate potential $\lambda_i$ with a rational $\alpha=p/q$ with $p$ and $q$ being co-prime integer.
$\alpha$ tunes the modulation period which can control the size of the primitive cell. In our system the size
of the primitive cell $\gamma$ is the least common multiple of $\{2,q\}$. The last term is the long-range interaction with a $1/r$-tails and
$i,j$ are not on-site. For the sake of simplicity, all the sawtooth in our model are chosen as regular triangle,
and the lattice constant between $2i$ and $2(i+1)$ is set unit, $t_1=1$ as the unit of energy and make $\alpha=1/q$.


\section{Result}
\begin{figure}[tbp]
\includegraphics[height=9cm,width=10cm] {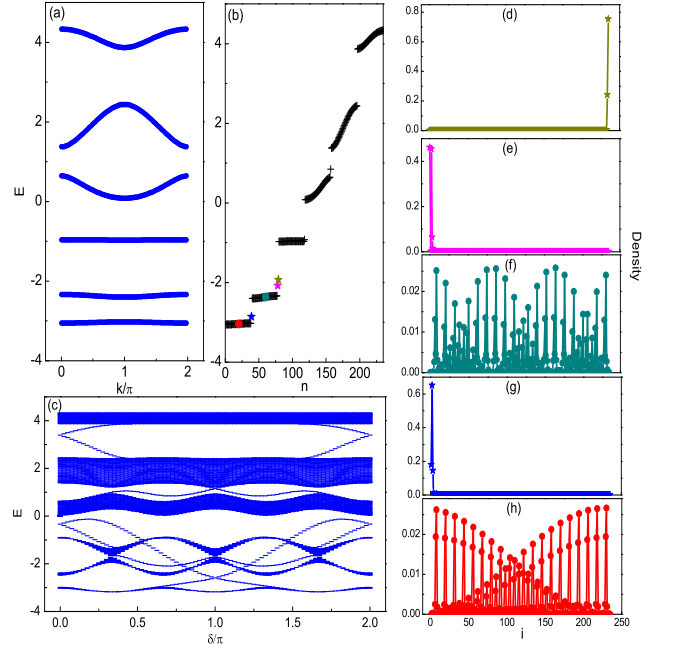}
\caption{(Color online) (a) Energy bands for the system with $N_{cell}=101$, $\delta=3\pi/4$
under PBC. (b) Energies in ascending order for the system with $235$ sites and $\delta=3\pi/4$
under OBC. (c) Energy spectrum varying with the phase $\delta$ with $L=235$ under OBC.
(d)-(h) Density distributions of the points signed by specific symbols in (b).
The edge states are signed by ''stars" and the states in bulk regime are mark by ''circles" in (b).
Here, $t_2=\sqrt{2}t_1,t_1=1,\lambda=1.5,\alpha=1/6$.
}\label{Fig1}
\end{figure}
First we consider the system without interaction terms, and the single-particle spectrum is split into $\gamma$ bands under PBC
where $\gamma$ is the least common multiple of $\{2,q\}$. To give a concrete example, 
we focus on the case with the parameter $\alpha=1/6$, as
it is the relatively simple system which can generate nearly flat band with nontrivial Chern number, while when
$\alpha=1/2$, the system is topologically trivial.
Fig.1(a) shows the band structure of the system with $t_2=\sqrt{2}t_1,t_1=1,\lambda=1.5,\delta=3\pi/4$ under PBC.
The two energy bands of the standard sawtooth chain split into six due to the modulation. The
lowest three bands are nearly flat while the upper three bands are dispersion. The bandwidth of the lowest band
is about $1/25$ of the gap between the lowest two bands. The flat parts are derived from the flat band of
the one without modulation and the upper ones are from the extended part of the original model.

For the lattices under OBC, the translation invariance is broken and momentum $k$ is no longer a good
quantum number. We rank the energy in ascending order shown in Fig.\ref{Fig1}(b). There emerge edge
modes in the gap regimes which is marked by ''star''. Those modes are localized at the boundaries while
the modes in the bulk regime are extended states.
Fig.\ref{Fig1}(d),(e),(g) show the density distributions localized at left or right boundaries are
corresponding to the three points from top to bottom labeled by ''star'' in Fig.\ref{Fig1}(b), respectively.
The states with extended distributions in Fig.\ref{Fig1}(f),(h) are corresponding to the modes pointed by
''circle'' in the second and the first band shown in Fig.\ref{Fig1}(b), respectively.
The energy spectrum changes periodically by rolling the phase $\delta$ from
$0$ to $2\pi$. The positions of the edge modes in the gaps vary with the change of the phase $\delta$.
Fig.\ref{Fig1}(c) shows the energy spectrum of the periodically modulating sawtooth chain with $\alpha=1/6$
as the function of phase $\delta$ under OBC. The edge states connect the neighbouring
bands in the gaps, which indicates the bulk states are topologically nontrivial\cite{Kane2,SCZhang,Hatsugai}.

\begin{figure}[tbp]
\includegraphics[height=9cm,width=9cm] {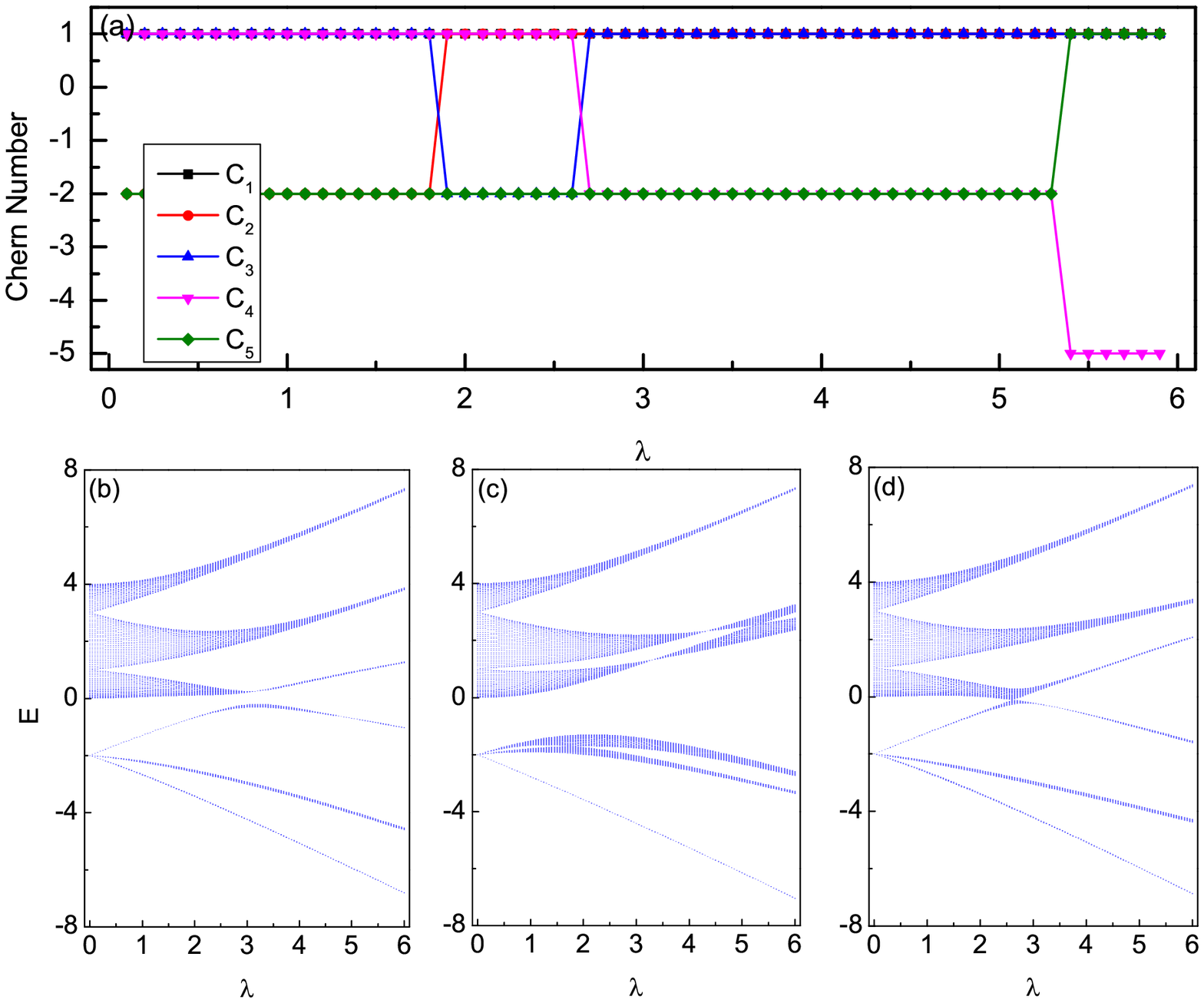}
\caption{(Color online) (a) The Chern number of lowest 5 bands as the function of the strength of the on-site
potential $\lambda$. The topological phase transitions accompanied topological indexes changing occur
at $\lambda_{c1}=1.80$, $\lambda_{c2}=2.61$ and $\lambda_{c3}=5.30$. (b)-(d) Energy varies as $\lambda$ with
$\delta=3\pi/4$, $\delta=0$ and $\delta=\pi/3$, respectively. (e)(f) are the enlarged views in (d) pointed
by boxes to show the gap closing clearly.  Here, $t_2=\sqrt{2}t_1,t_1=1,\alpha=1/6$.
}\label{Fig2}
\end{figure}
To characterize the topological properties of the systems, Chern number is a good candidate
\cite{Avron,Thouless,Berry,Simons,Fukui}. We define the Chern number assigned to the $n$th
band as $C_n$ in the parameter space of momentum $k$ and phase $\delta$, which is given by
$C_n=1/(2\pi i)\int \mathrm{d}k\mathrm{d}\delta [\partial_k A_{\delta}^n-\partial_{\delta}A_{k}^n]$,
where $A_{\mu}^n$ is the Berry connection, $A_{\mu}^n=i\langle \psi_n(k,\delta)|\partial_{\mu}|\psi_n(k,\delta)\rangle$
($\mu=k,\delta$), and $\psi_n(k,\delta)$ is the $n$th eigenstate in momentum $k$ and phase $\delta$ space.
And it is easy to find $\sum_{n=1}^{\gamma}C_n=0$, where $\gamma$ is the number of energy bands.
We get six $C_n$ corresponding to Fig.\ref{Fig1}(a) from bottom to top are $1,-2,1,1,-2,1$ where we choose $t_2=\sqrt{2}t_1,t_1=1,\lambda=1.5$.
With the change of $\lambda$, we find three topological phase transition points
$\lambda_{c1}\approx 1.80$, $\lambda_{c2}\approx 2.61$ and $\lambda_{c3}\approx 5.30$(see in Fig.\ref{Fig2}(a)).
According to Fig.\ref{Fig2}(a),the Chern number of the lowest band $C_1$ is always equal to unit. And when $\lambda<\lambda_{c1}$,
$C_2=-2$ and $C_3=1$. After crossing the first transition point $\lambda_{c1}$, $C_2$ and $C_3$ swap their values,
i.e. $C_2=1$ and $C_3=-2$. In the regime $\lambda_{c1}<\lambda\leq3$, $C_2$ keeps its value, $C_3$ steps to $1$ and $C_4$
changes from $1$ to $-2$. When $\lambda$ exceed $\lambda_{c3}$, $C_4$ turns to $-5$ and $C_5$ is from $-2$ to unit.

A topological phase transition accompanied with a change of topological numbers may
occur between two topological distinct phases. It has been widely known that the
energy gap close at the topological phase transition points. Another situation is found that the states
with different topological indexes can be continuously connected without gap closing, but with different symmetry\cite{DYXing,Nagaosa}.
For our model, it is former one, and we can understand the phase transition by the gap closing when
the parameter crossing the transition points. In Fig.\ref{Fig2}(b)-(d), we present the energy versus
the strength of modulation $\lambda$ with $t_2=\sqrt{2}t_1$, $t_1=1$, $\alpha=1/6$ and different $\delta$
under PBC. Fig.\ref{Fig2}(b) shows the case of $\delta=3\pi/4$. Along the whole parameter region, there
are no gaps closing. It seems there is no topological phase transition. By shifting the phase $\delta=\pi/3$(Fig.\ref{Fig2}(c)),
the gaps close at $\lambda\approx1.80$ and $5.30$ which are corresponding to $\lambda_{c1}$ and $\lambda_{c3}$.
In Fig.\ref{Fig2}(d), $\delta=0$ is chosen, and we can find the energy gap between the third and forth band closes at $2.61$
corresponding to $\lambda_{c2}$. In the following paper, we focus on $\delta=3\pi/4$, $\lambda=1.5$, $\alpha=1/6$ and the lowest band.




It has been demonstrate that our system with the subbands being fully filled by
fermions is a insulator with topologically nontrivial property. If the subband is partially filled, the system is no longer a
insulator but a topologically trivial conductor. Next we consider $N_f$ fermions are loaded into
the finite system with $N_{cell}$ primitive cells subjected to the long-range interaction
and the number of total sites is $L=qN_{cell}$ with the filling factor $\nu=N_f/N_{cell}$.
In following paper, we focus on the case of $\nu=1/3$, which is the lowest band being partially
filled. For the interaction term in the Eq.\ref{eqn2} with $\alpha=1/6$, we
make a truncation $|i-j|\leq 12$ where $i$, $j$ are the indexes of the lattice sites. When the
$i$th and the $j$th sites are at the same horizontal line, $|\vec{r}_i-\vec{r}_j|$ equals to min$\{|i-j|/2,(L-|i-j|)/2\}$.
For the case that they are at different horizontal lines, we set parameter
$l=(|i-j|-1)/2$, if $|i-j|\leq L/2$ and if $|i-j| > L/2$, $l=(L-|i-j|-1)/2$
and $|\vec{r}_i-\vec{r}_j|=\sqrt{l^2+l+1}$, where the number of sites $L$ is even.

\begin{figure}[tbp]
\includegraphics[height=8cm,width=9cm] {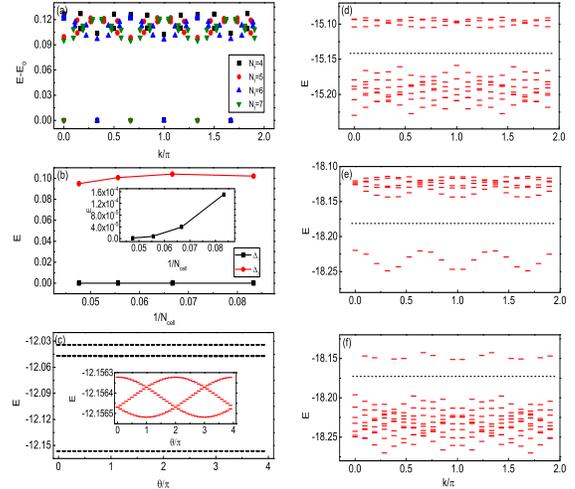}
\caption{(Color online) (a) Low-energy spectrum in momentum space for $N_f=4,5,6,7$ under PBC; (b) Finite-size scaling of
the bandwidths of the ground states $\Delta_1$ and the gaps $\Delta_2$. The subplot shows the bandwidths of the GS decrease
greatly as the scaling. (c) Spectrum flux versus phase factor $\theta$ for the system with $N_f=4$ under twist boundary condition.
The inset is the enlarged view of the spectrum flux of the lowest $3$ energy states. (d)-(f) Low-energy spectrum of the
quasihole excitations, (d) is for $N_f=5,N_{cell}=18$, (e) is for $N_f=6,N_{cell}=19$ and (f) is for $N_f=6,N_{cell}=21$.
Here, $t_2=\sqrt{2}t_1,t_1=1,\alpha=1/6$, $\lambda=1.5$ and $\delta=3\pi/4$.
}\label{Fig3}
\end{figure}

In the present of the interaction term, we diagonalize the Hamiltonian $H$ in each momentum space with $k=2\pi m/N_{cell}$
where $m$ is $0,1,\ldots,N_{cell}-1$. Due to the number of the bands being six for $\alpha=1/6$, it is difficult to
numerically study. However, in flat band limits, it can be diagonalized by projecting
onto the valence band and neglecting conduction bands, Which is similar to the method on dealing with
lowest-Landau-level(LLL) in FQHE\cite{Regnault1}.
For finite systems with a given filling factor, we study the low-energy spectrum changing
with different numbers of fermions. Fig.\ref{Fig3}(a) shows the low-energy spectrum of the system
in momentum sectors with the lowest bands being partially filled $\nu=1/3$,
$t_2=\sqrt{2}t_1$, $t_1=1$, $\lambda=1.5$, $\alpha=1/6$, $\delta=3\pi/4$ and different $N_f=4,5,6,7$.
For different $N_f$, it has $3$-fold degenerating ground states with a finite energy gap
separating the ground-state manifold from the low-energy excited states even with small
interactions due to the flat band limits, and the $3$-fold degenerate ground states emerge at momenta
$(2\pi)\{1/6,1/2,5/6\}$ for $N_f$ being even and $(2\pi)\{0,1/3,2/3\}$ for $N_f$ being odd. It is
similar to the usual fractional quantum hall effect on the torus, and we can set $N_{\Phi}=N_{cell}$, where
$N_{\Phi}$ is the number of the flux quanta in the single-particle orbits of the LLL. According to generalized
Pauli principle, no more than one particles can occupy $\eta$ consecutive orbits
for the filling factor $\nu=1/\eta$\cite{Regnault1,Nakamura}.
It is easy to find the lowest $\eta$ states emerge at the determinate positions in momentum space
$K=(2\pi)\{[\eta N_f(N_f-1)/2+\kappa N_f]\% N_{cell}\}/N_{cell}$ where $\kappa$ is $0,1,\ldots,\eta-1$ for the filling
factor $\nu=1/\eta$ and $\%$ means access module\cite{xuzhihao}.

In Fig.\ref{Fig3}(b), it shows the width of the lowest three states $\Delta_1=E_2-E_0$ and $\Delta_2=E_3-E_2$ is
the energy gap between the third state and the forth state versus $1/N_{cell}$. The splitting between the
$3$-fold ground states and the higher parts is finite and does not decrease which extrapolates to a finite
at large $N_{cell}$ limit. In the subplot of the Fig.\ref{Fig3}(b), the bandwidths $\Delta_1$ are greatly
decreased as the increasing of the scaling. The ground state splitting $\Delta_1$ is always significantly
smaller than the energy gap $\Delta_2$ for various system sizes.

For characterized topological properties of the many-body states, we chooses the twist boundary condition
$\psi(r_i+L,\delta)=e^{i\theta}\psi(r,\delta)$, where $\theta$ is the phase factor. Under the twist boundary condition,
each momentum $k$ takes a shift $k\rightarrow k+\theta/N_{cell}$, while the total momentum is still a good quantum number.
By varying phase factor $\theta$, the spectrum flux can be obtained which is shown in Fig.\ref{Fig3}(c) for the system with
$\nu=1/3$, $t_2=\sqrt{2}t_1$, $t_1=1$, $\lambda=1.5$, $\alpha=1/6$, $\delta=3\pi/4$ and $N_f=4$. It shows that the lowest $3$
degenerate states flow into each other and are separated by a obvious gap from the excited parts. From Fig.\ref{Fig3}(b),
it is easy to find the bandwidths $\Delta_1$ are extraordinary small comparing
with the gaps at large $N_f$. 
The inset of the Fig.\ref{Fig3}(c) is the enlarged view of the spectrum flux of the lowest $3$ energy states.
For the case of $\theta=0$ is corresponding to the system under PBC. The behavior of the flux indicates the unit Chern number
which is defined in $\{\theta,\delta\}$ parameter space and the numerical calculation follows \cite{Fukui}.


In order to investigate the fractional statistics of the fractional topological states,
the quasihole excitations are studied. The counting of the quasihole excitations is based
on the generalized Pauli principle\cite{Regnault1}. The number of quasihole excitations with the filling
factor $1/\eta$ can read as
\begin{equation}
\label{eqn3}
N_{qh}=\frac{N_f(N_{cell}-(\eta-1)N_f-1)!}{N_f!(N_{cell}-\eta N_f)!}.
\end{equation}
As shown in Fig.\ref{Fig3}(d), for the case of $N_f=5$ and $N_{cell}=18$ which is
removing one fermion from the system with $\nu=1/3$ filling, the quasihole spectrum shows a distinguishable
gap dividing the $126$ low-energy quasihole states from the higher part. In Fig.\ref{Fig3}(e),
we present $N_f=6$ and $N_{cell}=19$ by inserting a flux into the system with $\nu=1/3$. Below the gap,
the number of quasihole states is $19$.
Both cases in Fig.{\ref{Fig3}}(d),(e) are for the cases of the greatest common divisors(GCD) of $\{N_f,N_{cell}\}$ equal to $1$.
They have the same number of the different momentum sectors below the gaps, respectively. While in the Fig.\ref{Fig3}(f),
one particle is removed from the system with $\nu=1/3$ and $N_{cell}=21$ and the GCD of $\{N_f,N_{cell}\}$ is $3$. In such
case, the momenta $\mathrm{mod}\{m,3\}=0$ have $10$ states below the gaps, while
the other cases have $9$ states in quasihole subspace where $m=kN_{cell}/(2\pi)$ and $k$ is momentum.
And the total number of quasihole excitations is $194$. These cases
show the importance of the finite-size scaling in quasihole excitations.



\section{Summary}
In conclusion, we present flat-band with topologically nontrivial index in quasi-one-dimensional modulated
sawtooth chain. By varying the strength of the modulation, the system undergoes a series of topological
phase transitions.  Recently, a novel method to probe topological transitions using a continuous deformation
between two systems with distinct topological numbers. The subgap states localized within the deformation area
are observed as evidence of the phase transition\cite{Verbin}. We also demonstrate the existence of fractional
topological states in such model with long-range interactions, which are supported by topological degeneracy,
the unit total Chern number of the degenerate ground states and fractional statistic of the quasihole.
Thanks to the development of cold atomic technology\cite{Giorgini} and tunable gauge potential\cite{Struck,Sacha}, the experimental realization
of our model with normal hoppings is much more possible in ultracold atomic gases than previous cases.


\begin{acknowledgments}
\end{acknowledgments}
Z. Xu thanks Prof. S. Chen, Dr. Z. Liu, Dr. X. Cai and L.-J. Lang for the helpful discussions.

\end{document}